\documentstyle[aps,graphicx]{revtex}

\newcommand{\BEQ}{\begin{equation}}
\newcommand{\EEQ}{\end{equation}}
\newcommand{\BEA}{\begin{eqnarray}}
\newcommand{\EEA}{\end{eqnarray}}

\renewcommand{\d}{{\rm d }}
\newcommand{\p}{\partial}
\newcommand{\ext} {\mathop{\rm ext}}
\newcommand{\nn}{\nonumber }
\newcommand{\Tr}{{\rm Tr}}
\newcommand{\eps}{{\epsilon}}
\include{macro}

\begin{document}

\title{Complexity of the Sherrington-Kirkpatrick Model in the Annealed 
       Approximation}
\author{A. Crisanti, L. Leuzzi, G. Parisi and T. Rizzo\\
Dipartimento di Fisica, SMC and INFM, Universit\`{a} di
Roma ``La Sapienza'', 
\\
Piazzale Aldo Moro 2, I-00185 Roma, Italy}
\maketitle
\begin{abstract}
A careful critical analysis 
of the complexity, at the annealed level, 
of the Sherrington-Kirkpatrick model has been performed.
The complexity functional is proved to be always invariant under the 
Becchi-Rouet-Stora-Tyutin
supersymmetry, disregarding the formulation used to define it.
We consider two different 
saddle points of such functional, one satisfying the supersymmetry 
[A. Cavagna {\it et al.}, J. Phys. A {\bf 36} (2003) 1175] and the 
other one breaking it
[A.J. Bray and  M.A. Moore, J. Phys. C {\bf 13} (1980) L469].
We review the previews studies on the subject, linking 
different perspectives and pointing out some inadequacies
and even
inconsistencies in both solutions.
\end{abstract}
\date{\today}
\maketitle

  \section*{Introduction}
  \label{s:intro}
 
The organization of thermodynamic states in complex systems, fundamental in
the 
 understanding of  
the dynamic properties, is a rather difficult task to deal with.
The quenched disorder characterizing these systems and the consequent
 frustration are such that a huge number of stable and metastable states
arises, even growing exponentially with the number of elements composing
the system.
In order to throw light on the structure of the landscape of the thermodynamic
potential, a very important theoretical tool is the logarithm of the number 
of states, either called {\em complexity} or {\em configurational entropy}.

In structural glasses, seen as disordered amorphous solids and, thus, treated
by the techniques of complex systems, at the glass transition the entropy drops
 to a (often much smaller) value going from the liquid to the solid state
and the states, as opposed, e. g.,  to the crystal states
do not display any specific symmetry.
 The condensed system has lost the ability of visiting different
 states (at least on the observation time-scale   considered) and this 
codifies into the loss of entropy. 
All the other possible states, not selected at the moment of
the transition of the liquid to a glass, are, anyway, still there from a 
statistical point of view  and could, in principle, still be reached on much 
larger time-scales. The complexity counts the many equivalent
states
 that could have been chosen at the moment of the quench.

In disordered mean-field models for glasses, e.g. the $p$-spin interaction
spin glass models with stable one-step Replica Symmetry Breaking (1RSB)
frozen phase \cite{Derrida,Gardner,CS}, 
the presence of many metastable states can be detected looking at the 
relaxation that displays
a dynamical transition, with diverging relaxation time scales,
at a temperature (the {\em dynamical temperature})
above the point where a true thermodynamic transition 
takes place, called {\em static temperature}.
In the range between the static and the dynamic  temperature 
the complexity displayed turns out to be
extensive.

What happens moving from 1RSB-stable spin glass models
to spin glass mean-field 
models whose
frozen phase is 
described by means of a Full Replica Symmetry Breaking 
(FRSB) solution?

In the last twenty years basically  two different proposal
have been put forward for the nature and the behavior of the complexity
of the
spin-glass mean-field Sherrington-Kirkpatrick (SK) model \cite{SKPRL75}, 
the prototype of mean-field spin-glass models.
The first one was originally presented
by Bray and Moore  \cite{BMan}; there the complexity was analyses both in the 
{\em annealed} approximation, i.e. as the logarithm of
the disordered average of the number
of states, and as  quenched average, i.e. the average of the logarithm, 
performed making use of the replica trick
in the case of replica symmetry.
The second one was initially proposed in Ref. \cite{PP}, 
where Parisi and Potters showed that the  complexity could be
 obtained by calculating the partition function of
$m$ distinct real replicas of the system \cite{MPRL95}
and provided the connection with the previous formalism by means 
of the generalization of the {\em two-group Ansatz} \cite{BM2g}.
In that context the annealed solution was called 'unbroken' 
two-group. 

Over the years it has become more and more evident that an important role
in the study of the complexity of disordered systems is played by the so-called
Becchi-Rouet-Stora-Tyutin (BRST) symmetry. Such a property, first discovered
 in the quantization of gauge theories \cite{BRST} is a {\em super}symmetry, 
in the sense that it transforms bosons into fermions and vice-versa.
In the context of  stochastic field equations,
it can be shown that the  integration of the  generating functional 
of correlation functions over disorder
leads to an action presenting BRST symmetry \cite{ZZ} (for 
the random field Ising model case see also Refs. \cite{RFIM}).
The integrated generating functional  formally coincides
with  the average over the quenched random couplings
of the number of metastable states of mean-field spin-glass
models.
In such a context, the property of BRST-susy invariance has recently
been analyses in Refs.  \cite{CGG,CGPM}. 
In this approach,
imposing the invariance helps in simplifying the
computation  and it is equivalent to set the due interdependence
between  the physical objects deriving from the TAP free energy functional
and composing the action, namely the TAP functional itself, its first 
derivatives with respect to the average site magnetization (i.e. the TAP 
equations) and  its Hessian.
In Ref. \cite{CGPM} a BRST symmetric annealed complexities is built, 
while the BM  complexity
functional appears to violate such symmetry.

Generally speaking, the fact that 
the solutions of a set of equations are BRST-invariant is connected with 
some robustness of the equations under linear perturbation.
Let us take into account the equations $\p_{m_i} F=0$,
 where $F$ is the thermodynamic 
functional  and the derivative is taken with respect to the microscopic 
variable $m_i$, $i=1,\ldots,N$ and let us assume that
the functional computed over the solutions takes   
If we perturb the equations by means of small external fields,
$\p_{m_i} F=0 \to \p_{m_i} F=h_i$,  
the whole set of solutions could, in principle, drastically
change. In general, solutions will appear/disappear at different 
values of $F$ 
with no given prescription for the relative transformation of the 
complexity function and its dominion.
The BRST susy can be recovered assuming that the number of solutions
at a given value of $F$ does not change
In Sec. \ref{ss:BRST} we present an argument the explain
such correspondence.

 Both the approach of Ref. \cite{BMan} and the one of Ref. \cite{CGPM}
identify an extensive
 complexity, computing the number of solutions
of TAP equations \cite{TAP} 
 in the annealed approximation and with the further simplification of 
neglecting the modulus of the determinant of the Hessian of the 
TAP free energy functional. 
However, the two complexities display 
 many differences.
They have a
different lower band-edge value of the complexity curve vs. free energy and a
different dominion in free energy, with different thresholds.
Their magnitude is different, BM complexity being much larger than
the one satisfying the BRST-susy. 
Performing the integration over the whole interval of possible
free energy values yields a finite result in the BM case, but zero in the 
BRST-susy one\footnote{This would be the total 
number of solutions if the modulus
of the determinant were not neglected.}. 
At the threshold, the behavior is once again qualitatively different:
the BM complexity goes to zero smoothly, whereas the  BRST-susy one
 drops discontinuously to zero.

What do the two different complexities represent and which one of the two
 is the ``right'' one, yielding proper 
information over the organization of the states?
Moreover, what do these quantities become in an exact
FRSB quenched computation?

We study the complexity of the Sherrington-Kirkpatrick (SK)
model,
 mean-field spin model with quenched disorder,
critically 
reviewing the analysis made at the annealed level in the far and near past,
linking apparently different approaches and discussing the role of 
BRST supersymmetry in this contest. 
We will carefully look at the limitations
of the annealed approaches, in particular
 from the point of view of physical
stability, including the incompleteness of the one-step RSB
approximation for the SK model.

Our aim is to present here a comprehensive picture of the annealed level,
leaving for elsewhere
the study of the quenched (FRSB) complexity \cite{CLPR}.

Before dealing with details and derivations we, now,
very briefly anticipate 
the main results of this paper and we outline
the scheme of their presentation.

{\em The BM formalism} \cite{BMan,DDY} {\em  is equivalent to  
the  supersymmetric one} as
presented in Refs. \cite{CGPM,CGG}.
Both at the  'microscopic' level  (site commuting and anti-commuting
variables) and at the 'macroscopic'  level (order parameters)
the actions in the two formalisms are related to each other
by a simple change of variables.
The form of BRST transformations and of the
Ward identities are also obtained  in the BM notation.
We, thus, can consider just one (supersymmetric) functional 
generating one set of saddle point equations.

{\em Both the BRST-susy and the BM solution} (which we will often refer to
as BRST-susy-breaking solution)   for the annealed complexity
 {\em are solution
of the same set of saddle point equations}.
 
In Secs. \ref{ss:inva}, \ref{s:SPan} we derive such properties, 
comparing the two approaches in all essential details.

Afterwards, 
we give some hints in order to understand the physical meaning of the
BRST-susy, showing, in Sec. \ref{ss:BRST}, the connection 
between such property and the property of {\em non-bifurcation} of
TAP-solutions of linearly perturbed TAP-equations with respect
to the unperturbed solutions and the BRST-susy.
In  Sec. \ref{s:1rsb}
the identification between the annealed BRST-susy solution and the static
 1RSB solution at zero magnetic field,
main topic of Ref. \cite{CGPM},
 is rederived in the case of the BM formulation. 
Moreover, in Sec. \ref{s:2G}, 
we recast everything in the formulation of the generalized 
{\em two-group Ansatz} of Ref. \cite{PP} and we show that breaking
 the supersymmetry amounts to consider a
non-trivial Ansatz
 in the replica calculation.  

In the second part of the paper, section \ref{s:stab_anneal},
we discuss 
several issues connected with the problem of
selecting the right solution between the BM and the BRST supersymmetric one.
In particular, we show that {\em the sign of the determinant of the Hessian
cannot be determined by the saddle point 
solution at the leading order in $N$}
and, applying Plefka's criterion to the 
analysis of the physical meaning of TAP solutions,
we explain why the parameter 
$B$ entering the determinant has to be set equal to zero, as in Refs.
\cite{BMan,PP,CGG,CGPM}, 
and why, as opposed to what is stated in Refs. \cite{BMan,commentBM},
such a choice {\em does not guarantee the positivity of the determinant} of 
the Hessian of the TAP free energy functional.

We also recall the results of Kurchan \cite{K91}
about the spontaneous BRST-susy breaking taking place on the BM saddle point:
the BRST-susy is recovered
by the analysis of the prefactor of the exponent of $N\Sigma_{BM}$ 
showing it to be zero on the BM saddle point.
 This result, while ensuring mathematical consistency,
 poses the problem of which is the correct prefactor of the saddle point 
when the modulus of the determinant is not dropped.

As a byproduct,
dealing with the problem that the true thermodynamic description
of the low temperature phase of the SK model is yielded by a FRSB
solution instead of a 1RSB one, in Sec. \ref{ss:bmy} we show that
 {\em the complexity proposed
by Bray, Moore and Young} \cite{BMY}, as quenched analogue of the
above broadly mentioned annealed complexity, {\em is computed over
  a BRST-susy saddle point}.


  \section{Counting the TAP-states:  the 
             complexity in the annealed approximation and the BRST-susy.}
  \label{s:annealed}
Before  discussing the properties of the number of solutions 
of the mean-field equations for the SK model,
we very shortly summarize the basic, widely known, features of
the Thouless-Anderson-Palmer (TAP) \cite{TAP} formulation.  
The TAP equations for the local average magnetization $m_i$ of the $i$th spin
are 
\BEA
&&m_i=\tanh\left[\beta \left(\tilde{h}_i-\beta (1-q)m_i\right)\right]
\ \ \ \ \forall i = 1,\ldots, N
\label{TAP_eq}
\\
&&\tilde{h}_i=\sum_j J_{ij} m_j
\label{h_tilde}
\EEA
where $q$ is the self-overlap of TAP configurations:
\BEQ
q\equiv \frac{1}{N}\sum_im_i^2
\EEQ
and $J_{ij}$ are distributed as 
\BEQ
P(J_{ij})=\sqrt{\frac{N}{2\pi}}\exp\left(-N\frac{J_{ij}^2}{2}\right)
\EEQ
The associated free energy functional is
\BEA
&&F_{\rm tap}(\{m\}) = E_{\rm tap}(\{m\}) - T S_{\rm tap}(\{m\})
\label{TAP_fen}
\\
&&E_{\rm tap}(\{m\}) = 
-\sum_{i<j}J_{ij} m_i m_j
-\frac{N\beta}{2}\left(1-q\right)^2
\label{TAP_ene}
\\
&&S _{\rm tap}
=\sum_i s_{\rm tap}(m_i;q)
=\sum_i\left\{ 
\log 2 -\frac{\beta^2}{4}(1-q)^2
-\frac{1}{2}\log\left(1-m_i^2\right)+m_i\tanh^{-1}m_i
\right\}
\label{TAP_entro}
\EEA
where $E_{\rm tap}$ and $S_{\rm tap}$
 are, respectively, the expressions for the internal energy and
 the entropy of a $\{m_i\}$ configuration and Eq. (\ref{TAP_eq}) is yielded by 
differentiating Eq. (\ref{TAP_fen}) \cite{TAP}: 
$\partial_{m_i} F_{\rm tap}= 0$. Furthermore one  defines
the disorder-independent ``field''
\BEQ
g(m_i;q)\equiv \frac{1}{\beta}\tanh^{-1}m_i+\beta(1-q) m_i
\EEQ
so that Eq. (\ref{TAP_eq}) can be reformulated as
\BEQ
\partial_{m_i} F_{\rm tap}=g(m_i;q) -\sum_j J_{ij} m_j=0
\label{TAP_eq2}
\EEQ

Using Eq. (\ref{TAP_eq2}), the expression for the energy of a TAP solution
$\{m_i^{\rm sol}\}$ can be written without making use
of the disorder $J_{ij}$ as 
\BEQ
E(\{m^{\rm sol}\})
= -\frac{1}{2\beta}\sum_{i}g(m_i;q)~m_i-\frac{\beta}{2}\left(1-q\right)^2
=-\frac{1}{2\beta}\sum_i m_i\tanh^{-1} m_i 
- \frac{\beta}{2}\left(1-q^2\right)
\label{TAP_ene2}
\EEQ

Combining this last result with the expression (\ref{TAP_entro})
 for the entropy one is 
able to formally rewrite the TAP free energy as a sum of single 
site-free energies of TAP solution, as
\BEQ
F(\{m^{\rm sol}\})=\sum_{i}f(m_i;q)
=\frac{1}{N}\sum_i\left\{-\log 2 -\frac{\beta^2}{4}(1-q^2)+\frac{1}{2}
\left[m_i \tanh^{-1} m_i+\log\left(1-m_i^2\right)\right]
\right\}
\label{fen_TAP}
\EEQ
where $f(m_i,q)$ is the site free energy computed on a solution of Eq. 
(\ref{TAP_eq2})

\subsection{The complexity in the annealed approximation} 

What we are interested in, is  the number of solutions of the TAP equations
at different free energy levels, which can provide substantial information
in order to understand the free energy landscape and, therefore, 
the organization of the states. If we label each of the
$\cal{N}$ solutions 
with the index $\alpha = 1, \ldots, \cal{N}$,
the number of solutions having free energy density equal to $f$ is 
\BEQ
\rho_s(f)=\sum_{\alpha=1}^{\cal{N}}
\delta\left[F_{\rm tap}(\{m^{\alpha}\}) - N f\right]
\EEQ
This can be formally transformed as
\BEA
&&\rho_s(f)=\sum_{\alpha=1}^{\cal{N}}\int_{-1}^{1}\prod_{i=1}^N \d m_i~
 \delta\left(m_i-m_i^{\alpha}\right) 
\delta\left[F_{\rm tap}(\{m\}) - N f\right]
\\
&&=\int_{-1}^{1}\prod_{i=1}^N \d m_i~
\delta\left(\partial_{m_i} F_{\rm tap}(\{m\})\right)
\left|{\rm det}(\partial_{m_i}\partial_{m_j} F_{\rm tap}(\{m\})\right|
\delta\left[F_{\rm tap}(\{m\}) - Nf\right]
\EEA
where the Hessian takes the form
\BEQ
\partial_{m_i}\partial_{m_j} F_{\rm tap}(\{m\})=
-J_{ij} + \left[\frac{1}{\beta}\frac{1}{1-m_i^2}+\beta(1-q)\right]\delta_{ij}
+O\left(\frac{1}{N}\right).
\label{Hessian}
\EEQ
Terms of order $1/N$ will be neglected since they are not relevant for the
present discussion.

At this point,  one can compute the 
annealed complexity, i.e. the logarithm of the average over the disorder of the
density of TAP solutions: 
\BEQ
\Sigma_a = \frac{1}{N}\log {\overline{\rho_s(f)}}.
\label{S_a} 
\EEQ
where the overline represents the average over the distribution
of the $J_{ij}$'s.
Details of such computation for  the SK model 
can be found both in the original paper of
Bray and Moore (Ref.  \cite{BMan}) 
and in the supersymmetric formulation of Ref. \cite{CGPM}, as well as 
in many other works, e.g. Refs. \cite{DDY,TanaEd,DDGGO}.
Here, we only stress the basic fact 
that, in both procedures, the determinant of the Hessian is 
taken without absolute value, thus implying that the quantity
actually computed would coincide with the ``true'' complexity (even though
in the annealed approximation) only if the determinant of all
 the solutions we are counting were positive.
  Such an assumption
is hard to verify in the SK model.
 

\subsection{BRST-invariance of the density of TAP-solutions.}
\label{ss:inva}
The density of TAP solutions can be written making use both of anti-commuting 
$\{\psi\},\{{\overline{\psi}}\}$
and commuting variables $\{m\},\{x\}$ as \cite{CGG,CGPM}
\BEA
&&\rho_s(f)= 
\int {\cal D}x~{\cal D}m~{\cal D}~{\overline \psi}~{\cal D}\psi~
e^{\beta {\cal S}(m,x,{\overline \psi},\psi)}
\\
&&{\cal S}(m,x,{\overline \psi},\psi)=
\sum_ix_i \p_{m_i} F_{\rm tap}(\{m\}) 
+\sum_{ij}{\overline\psi}_i\psi_j\p_{m_i}\p_{m_j}F_{\rm tap}(\{m\})
+u\left[F_{\rm tap}(\{m\})-Nf\right]
\label{actionpsi}
\EEA
where ${\cal D} a\equiv \mbox{prefactor} \times \prod_i \d a_i$.
The transformation under which such an action is invariant
is \cite{ZZ,CGG,CGPM}:

\BEQ
\delta m_i=\epsilon ~\psi_i
\hspace*{1 cm}
\delta x_i=\epsilon~ u~\psi_i
\hspace*{1 cm}
\delta{\overline\psi}_i=-\epsilon~x_i
\hspace*{1 cm}
\delta\psi_i=0
\label{BRST_fields}
\EEQ

The Ward identities generated by imposing the invariance 
with respect to the above transformations
of 
observables
 are, e.g.,
\BEA
&&\left<{\overline\psi}_i\psi_i\right>=-\left<m_ix_i\right>
\label{W1}
\\
&&u\left<{\overline\psi}_i\psi_i\right>=\left<x_i^2\right>
\label{W2}
\EEA
where the average $\left<\left(\ldots\right)\right>$
 is computed over the measure $e^{\cal S}$.

Bray and Moore \cite{BMan} used explicitly, 
in their computation, the TAP equations (\ref{TAP_eq2}) in order
to simplify the procedure.
In Ref. \cite{CGPM} it was claimed 
that such a substitution brought to an action
no longer BRST invariant.
Such an inconvenient is, however, only apparent. Indeed, shifting 
the integration variables $\{x\}$ the BRST invariant form is readily
restored.
The action yielded by the BM procedure is different from the so called
susy one because of the insertion of Eq. (\ref{TAP_eq2}), which means
\BEQ
{\cal S}(x,m,{\overline\psi},\psi)={\cal S}_{\rm BM}(x,m,{\overline\psi},\psi)
+u\left[\frac{1}{2}\sum_{ij}J_{ij}m_i m_j-\frac{1}{2\beta}\sum_i
g(m_i;q)m_i\right]
\EEQ

Since the quenched disorder $J_{ij}$ enters the action as
$J_{ij}\left(x_im_j+{\overline\psi}_i\psi_j+u/2~m_i m_j\right)$, a shift
on the integration variable
\BEQ
x_i \to x_i'=x_i+\frac{u}{2}m_i
\EEQ
is enough to make both actions coincide:
\BEQ
{\cal S}(x,m,\psi,{\overline\psi})
={\cal S}_{\rm BM}(x+\frac{u}{2}m ,m,\psi,{\overline\psi})
\EEQ
In the new set of variables $\{x',m,\psi,{\overline\psi}\}$
the transformation (\ref{BRST_fields}) keeping a BSRT-susy function
invariant remains unchanged.
The shift of $x_i$ is performed over an integration variable, 
thus without affecting the final result, yet the transformations given in Eq.
(\ref{BRST_fields}) and the Ward 
identities computed with the BM measure 
in the original set of variables are changed.

In the old set of variables the BRST transformation read (as discussed also 
in Ref.  \cite{Alessia})
\BEQ
\delta m_i=\epsilon ~\psi_i
\hspace*{1 cm}
\delta x_i=\epsilon~ \frac{u}{2}~\psi_i
\hspace*{1 cm}
\delta{\overline\psi}_i=-\epsilon~x_i
\hspace*{1 cm}
\delta\psi_i=0
\label{BRST_fields2}
\EEQ
and 
the averages over the two actions are connected by
\BEQ
\left<a(x)\right> = \left<a\left(x-\frac{u}{2}m\right)\right>_{\rm BM}
\EEQ
where the average 
$\left<\left(\ldots\right)\right>_{\rm BM}$ is computed over the 
measure $e^{{\cal S}_{\rm BM}}$
so that Ward identities computed with the BM action become
\BEA
&&\left<{\overline {\psi}}_i \psi_i\right>_{\rm BM}
=-\left<m_ix_i\right>_{\rm BM}+\frac{u}{2} q
\label{Ward1bm}
\\
&&u\left<{\overline\psi}_i\psi_i\right>_{\rm BM}
=\left<x_i^2\right>_{\rm BM}- u \left<m_i x_i\right>_{\rm BM}+\frac{u^2}{4}q
\label{Ward2bm}
\EEA
Inserting Eq. (\ref{Ward1bm}) in Eq. (\ref{Ward2bm}) one gets
\BEQ
\left<x_i^2\right>_{\rm BM}=\frac{u^2}{4}q
\EEQ

Even if in the notation of Ref. \cite{CGPM}
it seemed that the action $S_{\rm BM}$ was not satisfying the BRST relations,
this was exclusively due to the fact that such relations
in the BM notation
read differently.

Moving to the macroscopic level, where the average number of solution is 
expressed as a function of the parameters $q$, $\Delta$, $\lambda$ and $B$,
the two approaches continue to be linked and equivalent to each other. 
The final expressions for the complexity in Refs. \cite{BMan} (BM)
and  \cite{CGPM} (CGPM) are:
\BEA
\mbox{BM functional}   &&
{\overline{\rho_s(f; u, B ,q, \Delta,\lambda )}}=
\ext_{\{u~ B~q~\Delta~\lambda\}}
~ \exp\left\{-\lambda~ q- \beta~u~f-(B+\Delta)(1-q)\right.
\label{rho_BM}
\\
\nn
&&\hspace*{8 cm}\left.
+\frac{(B^2-\Delta^2)}{2 \beta^2}+\log I \right\}
\\
&&I = \int_{-1}^1\frac{ \d m}{\sqrt{ 2 \pi \beta^2 q}}~ 
\left(\frac{1}{1-m^2}+B\right)
\exp\left\{-\frac{\left(\Delta~m-\tanh^{-1}m\right)^2}{2 \beta^2 q}
+\lambda~m^2
+u~f(m; q)\right\}
\\
&& f(m; q)= -\log 2-\frac{\beta^2}{4}(1-q^2)+\frac{1}{2}m\tanh^{-1} m
+\frac{1}{2}\log (1-m^2)
\label{fen_BM}
\\
\mbox{CGPM functional} && 
{\overline{\rho_s(f; u,  B,q,  \Delta,\lambda)}}=
\ext_{\{u~B~q~\Delta~\lambda\}}
 ~\exp\left\{
-\lambda~ q- \beta~u~f-(B+\Delta)(1-q)\right.
\label{rho_CGPM}
\\
&&\hspace*{4 cm}
\left.+\frac{(B^2-\Delta^2)}{2 \beta^2}-\frac{\beta^2}{4}u^2 q^2
- \Delta ~u ~q-\beta^2u~q(1-q)+\log I\right\}
\nn
\\
&&I= \int_{-1}^1\frac{ \d m}{\sqrt{2 \pi \beta^2 q}}~ 
\left(\frac{1}{1-m^2}+B\right)
\exp\left\{-\frac{\left(\Delta~m-\tanh^{-1}m\right)^2}{2 \beta^2 q}
+\lambda~m^2
+u~\Phi_0(m; q)\right\}
\\
&&
\Phi_0(m; q) =  -\log 2-\frac{\beta^2}{4}(1-q)^2
+m\tanh^{-1} m
+\frac{1}{2}\log (1-m^2)
\label{fen_CGPM}
\EEA

The first expression is 
obtained making explicit use of Eq. (\ref{TAP_ene2}) for the TAP energy,
while the  second one is computed without ever using
such a relation.
Apparently the two expressions  differ. 
Since, however, both actions describe the evolution in the parameter-space
of the same system, the two formulations must coincide
(exactly as the 'microscopic' description and the final 
results do)  introducing a proper change of variables.
Indeed,
to link  the two approaches the following transformation can be set:
\BEA
&&\Delta_{BM}=\Delta_{CGPM}+\frac{\beta^2}{2}u~q
\label{Delta_ch}
\\
&&\lambda_{BM}=\lambda_{CGPM}+\frac{1}{2}u ~\Delta_{CGPM} 
+\frac{\beta^2}{8}u^2 q 
\label{lambda_ch}
\EEA 
Notice that the two   functions 
$f(m; q)$ (in the BM formalism) and $\Phi_0(m; q)$
(in the formalism of Ref. \cite{CGPM})
are related to each other by
\BEQ 
f(m; q)=\Phi_0(m; q)-\frac{1}{2} m \tanh^{-1}m
-\frac{\beta^2}{2}q(1-q)
\label{Fen_ch}
\EEQ

Since the formulations are equivalent, the notation 
to adopt is no more important.
We will use the original notation of Ref. \cite{BMan},
not simply because it is the eldest one, but rather
to have a more direct comparison with other works, e.g. the one of
  Parisi and Potters \cite{PP},
that showed that the BM action can be 
obtained from the Legendre transform approach of Ref. \cite{MPRL95}  making a 
two-group Ansatz on the matrix $Q_{ab}$ entering the computation of the 
free energy of the coupled replicas (in this way they were able to find new 
solutions of the BM equations called 'unbroken two-group'), 
 or the quenched computation of the complexity of Refs. 
\cite{BMque,BMY}.
For simplicity we drop the subscript BM in the following.

\subsection{Saddle point equations}
\label{s:SPan}

The variational equations, yielding the saddle point(s) values of the 
parameters for the annealed 
complexity take the form:
\BEA
\frac{\p \Sigma_a}{\p u}=0 \hspace*{ 1 cm}&\to&
f = \left<f(m; q)\right>
\label{SP_f}
\\
\frac{\p \Sigma_a}{\p B}=0\hspace*{ 1 cm}&\to&
B=\beta^2\left(1-q-\left<\frac{1-m^2}{1+B(1-m^2)}\right>\right)
\label{SP_B}
\\
\frac{\p \Sigma_a}{\p q}=0\hspace*{ 1 cm}
&\to&
\lambda=B+\Delta-\frac{1}{2 q}
+\frac{\left<\left(\Delta~m-\tanh^{-1}m\right)^2\right>}{2\beta^2 q^2}
+\frac{\beta^2}{2}u~q
\label{SP_l}
\\
\frac{\p \Sigma_a}{\p \Delta}=0\hspace*{ 1 cm}
&\to&
\Delta=-\frac{\beta^2}{2}(1-q)+\frac{1}{2q}\left<m\tanh^{-1}m\right>
\label{SP_D}
\\
\frac{\p \Sigma_a}{\p \lambda}=0\hspace*{ 1 cm}&\to&
q=\left<m^2\right>
\label{SP_q}
\EEA
where the average
\BEA
&&\left<O(m)\right>=
\frac{1}{I}\int_{-1}^1 \d m ~O(m)~e^{{\cal{L}}(m; u, q, \Delta, \lambda)}
\label{ave}
\\
&&I=\int_{-1}^1 \d m ~e^{{\cal L}(m; u, q, \Delta, \lambda)}
\EEA
is taken over the action
\BEQ
{\cal L}(m; u, q, \Delta, \lambda)=\log\left(\frac{1}{1-m^2}+B\right)
-\frac{1}{2}\log\left(\beta^2 2 \pi q\right)
-\frac{\left(\Delta~m-\tanh^{-1}m\right)^2}{2 \beta^2 q}+\lambda~m^2
+u~f(m;q)
\label{action}
\EEQ
As noted in Ref. \cite{CGPM} the last term of Eq. (\ref{SP_l}) was missing 
in Ref. \cite{BMan}. The free energy of the TAP solutions 
$f(m;q)$ is the one expressed
in Eq. (\ref{fen_TAP}) or (\ref{fen_BM}).

Fixing $u$, therefore leaving $f$ as a free parameter,
the above equations have at least two different  solutions
\footnote{In practice, when we perform the numerical resolution  
of Eqs. (\ref{SP_l})-(\ref{SP_q})
we can see  that there are two solutions
of the saddle point equations  to which our computer  program converge and 
the selection of one solution instead
of the other only depends on the initial  conditions we give. }. 
One solution satisfies the two  relations of BRST supersymmetry as
stated in Refs. \cite{CGPM,CGG}, which we rewrite here in the present 
notation as 
\BEA
&&B+\Delta=-\frac{\beta^2}{2}u~q
\label{BRST1}
\\
&&\lambda=\frac{\beta^2}{8}u^2 q
\label{BRST2}
\EEA

The saddle point Eq. (\ref{SP_B}), 
substituting Eq. (\ref{SP_q}) in it, admits  solution for  
$B\geq0$ (see Appendix A for details).
In Sec. \ref{ss:ple} we recall  that a general criterion, 
formulated by Plefka \cite{Ple0},
 can be applied as a {\em necessary} condition
to select  physically relevant solutions.
We anticipate that this criterion requires $B=0$,
so that Eq. (\ref{BRST1}) becomes a condition for $\Delta$ alone.

\begin{figure}[th!]
\begin{center}
{
\includegraphics*[width=0.49\textwidth]{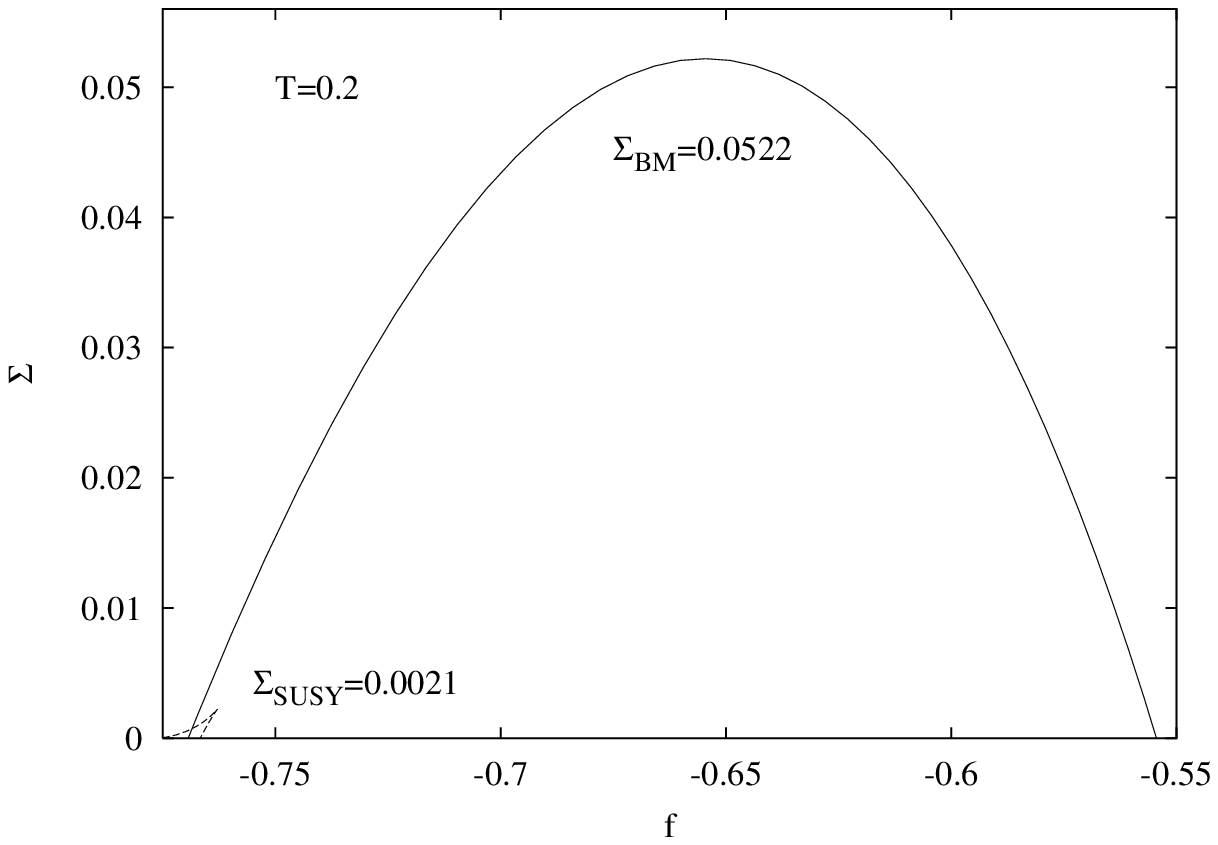}
\includegraphics*[width=0.49\textwidth]{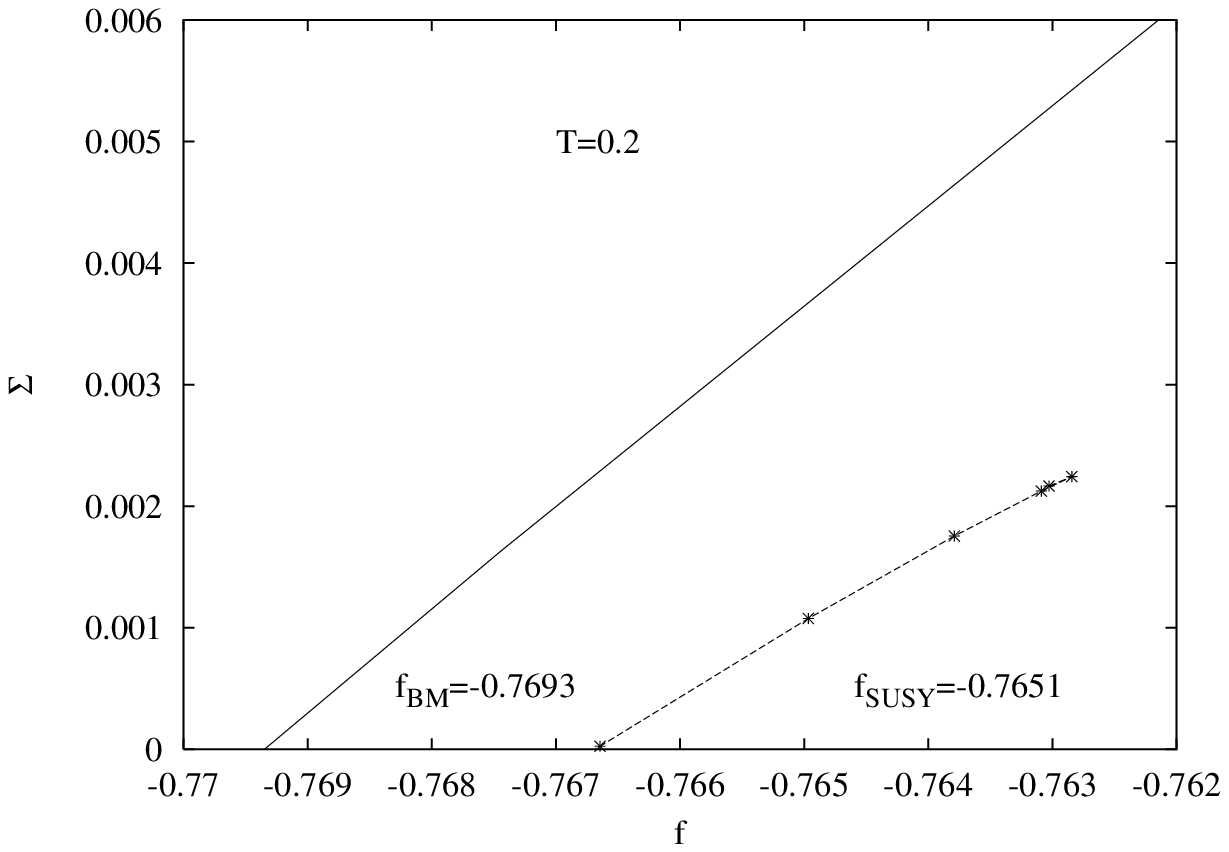}
}
\caption{Comparison of  complexities yielded respectively  by the BRST-susy 
solution (dashed line) and the BRST-susy-breaking solution (continuous line)
of the annealed saddle point equations at $T=0.2$.
The range of free energy values of the solution breaking the supersymmetry
of the TAP-action (\ref{action}) is much larger than the $f$-range 
of the susy one. The maximum complexity for this solution is also much
higher than the one for the BRST-invariant solution (the value at the cusp
of the dashed curve). Such a difference increases by lowering $T$.
}
\label{f:1}
\caption{Detail of the complexity curves for low free energy values.
The asterisks stand for the complexity of the static one-step RSB solution
at zero magnetic field, 
coinciding with the BRST-susy saddle point.
The values of free energy at which the complexity reaches zero 
(sub-exponential  growth of solutions of TAP equations with 
increasing size) are $f_{\rm BM}=-0.7693$, $f_{\rm susy}=-0.7651$,
both below the true equilibrium value at $T=0.2$:
$f_{\rm eq}=-0.7594$ (as computed, e.g.,  in the FRSB scheme).
Notice that the branch on the left-hand side of the BRST-susy solution
in not shown.}
\label{f:2}
\end{center}
\end{figure}

\begin{figure}[th!]
\begin{center}
{
\includegraphics*[width=0.49\textwidth]{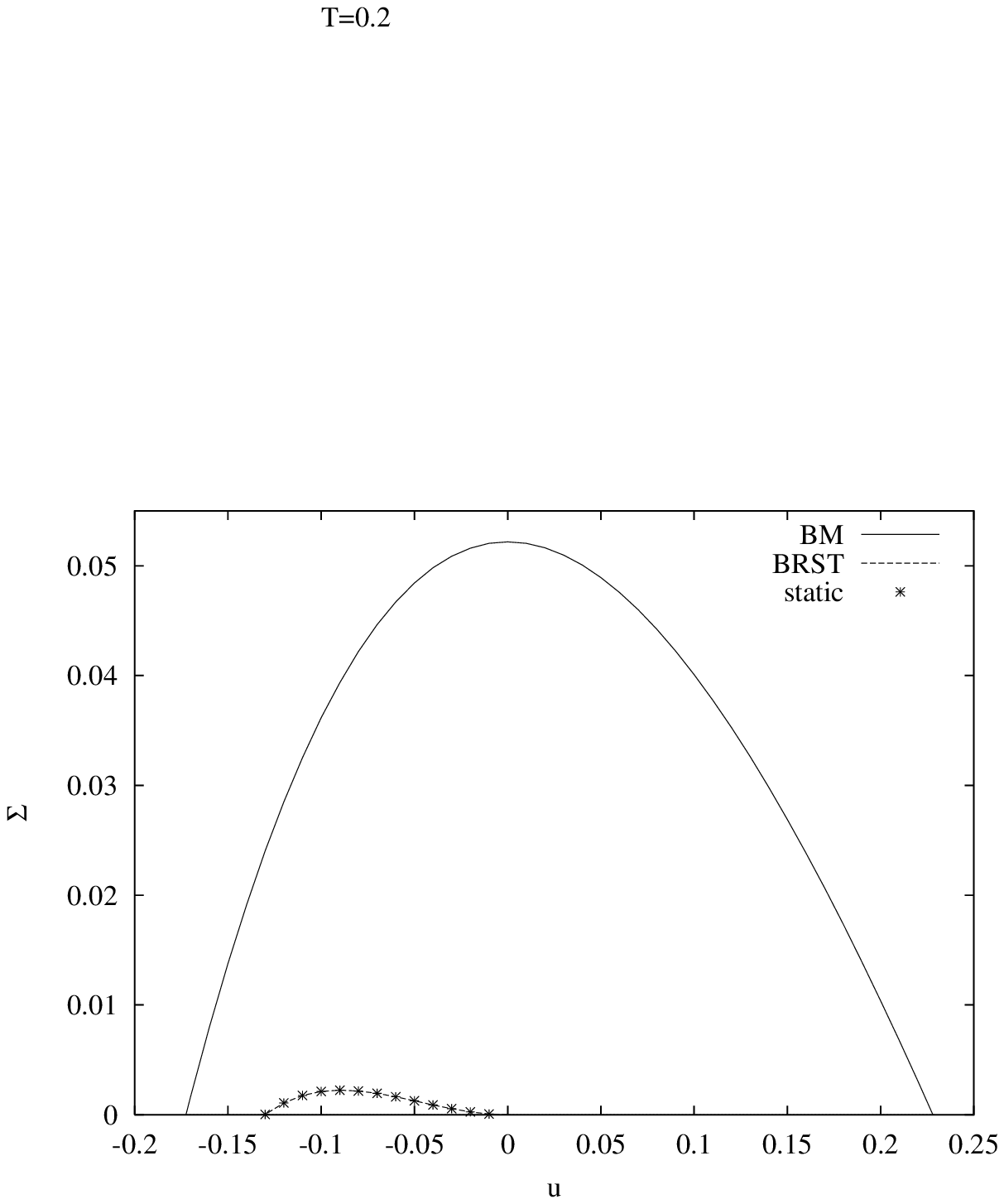}
\includegraphics*[width=0.49\textwidth]{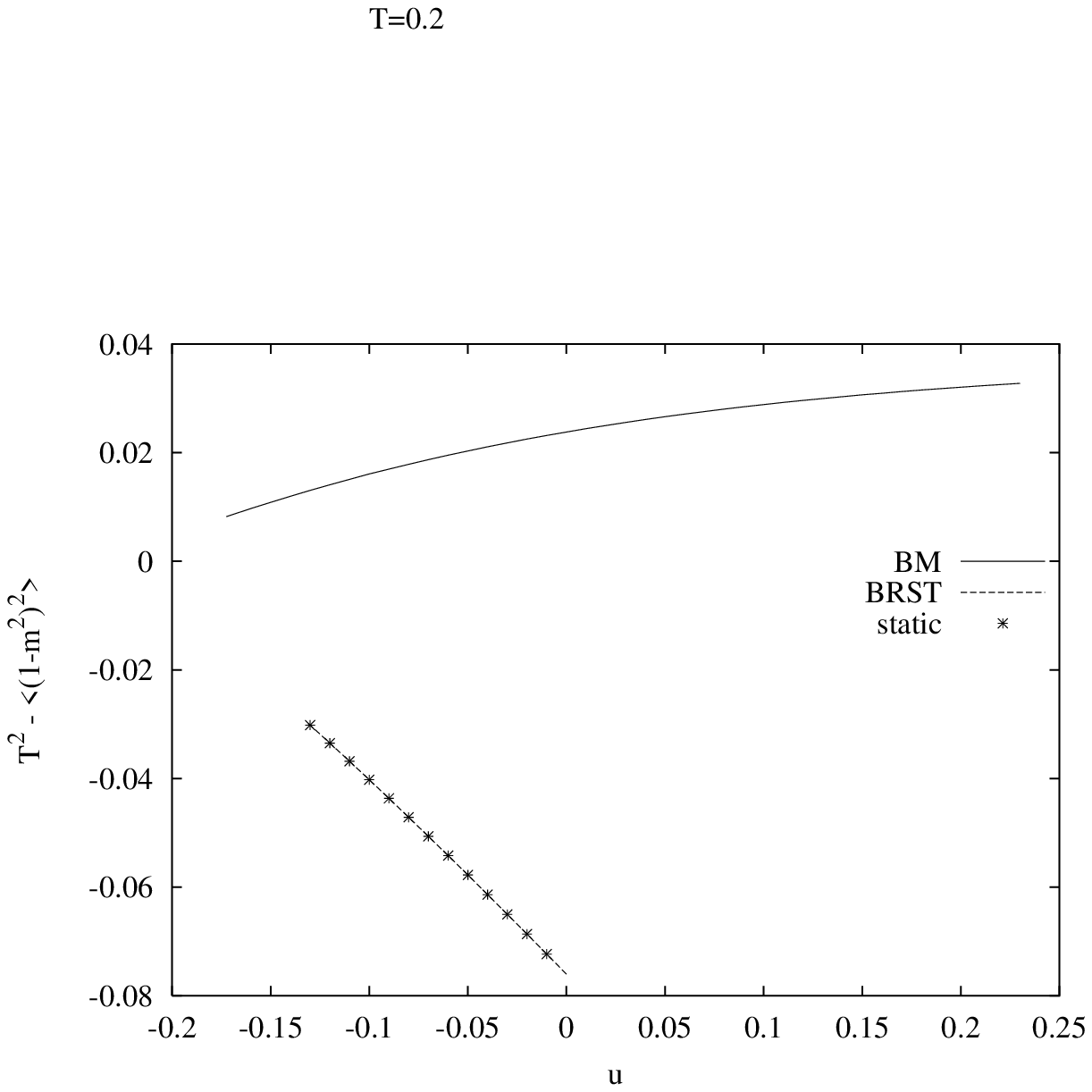}
}
\caption{Complexity vs. $u$ (conjugated to $f$) for both saddle points.
The BRST-susy complexity displays a maximum below zero (corresponding to the 
cusp in the $f$ dominion (see Fig. \ref{f:1})). It goes to zero as $u\to 0$.
On the contrary the solution with broken susy has a maximum for $u=0$.
}
\label{f:3}
\caption{The quantity
$T^2-\left<(1-m^2)^2\right>$ is plotted vs. $u$. 
Following the Plefka's criterion, if such a quantity is positive 
the solution is physically meaningful (see text for details), otherwise
it is not.
In the case of the BRST solution, it coincides with the replicon 
(stability eigenvalue of the $q_1$ fluctuations in the one-step RSB 
static solution with no external field), which we plot with asterisks.
From the plot one can see that Plefka's criterion is satisfied
only by the BM solution.
This means that if the annealed 
approximation in the complexity computation would be a reliable one, 
and if no further inconsistency would arise, the 
physically relevant saddle point over which it should be computed would
be the one breaking the BRST-susy.
}
\label{f:4}
\end{center}
\end{figure}

In Figures 
\ref{f:1}, \ref{f:2} and  \ref{f:3} we show the behavior, at $T=0.2$,
of both solutions, both versus $f$ and $u$.
The annealed complexity computed over the supersymmetric solution goes to zero
smoothly  as $u \to 0^-$, it displays a maximum at some
 $u_{\rm max}$ (or $f_{\rm th}$, if the
 behavior vs.  free energy is considered) and crosses the $u$-ax
at some $u_0$ such that $u_0<u_{\rm max}<0$.
Unlike $\Sigma_a(u)$, the curve $\Sigma_a(f)$ is not univocal: 
it displays
a cusp at $f_{\rm th}$ (see Fig. \ref{f:1}) and then turns back.

The other solution of the saddle point equations, which will refer to
as BM or BRST-susy-breaking solution,
is the one introduced in Ref. \cite{BMan}.
Such a solution does not satisfy relations (\ref{BRST1}-\ref{BRST2}), 
 thus spontaneously breaking BRST-susy,
 its dominion in $u$ (or $f$) is broader than the one of the symmetric
 solution, the maximum value of $\Sigma_a$ is for $u=0$ and, 
at any temperature,
it is larger than the correspondent value on the BRST-invariant solution.
The behavior of the annealed complexity is univocal
both in $u$ and in $f$.

\subsection{Physical Meaning of BRST-susy in the context of 
TAP-solutions.} 
\label{ss:BRST}

In this section we will follow a reasonment allowing for an intuitive
explanation of the physical meaning of the BRST supersymmetry
in terms of a particular behavior of the solutions of the TAP equations.

The objects of interest of the present approach are sums over TAP states,
multiplied by some weight function, typically of the form
$e^{\beta u  F(m)}$.
We make the hypothesis  that the set  of solutions over 
which we perform the sums
does not qualitatively change 
 adding a small magnetic field to a single spin $k$.
This means
 that each solution $\alpha$ with associated magnetizations $\{m_i^{\alpha}\}$ 
goes continuously to a new solution in presence of the field
 $h_k$, defining the  functions $\{m_i^{\alpha}(h_k)\}$.

Given a generic observable depending 
on the magnetizations, its white average, i.e. the average value that
the 
observable $g$ takes over each TAP solution, is given by:
\begin{equation}
\left[ g(m)\right]_{\rm sol}\equiv
\frac{1}{{\cal N}_s}\sum_{\alpha}g(m^{\alpha}_i)
\end{equation} 
where ${\cal N}_s$ is the number of solutions we are summing over.

We, now, assume that  
(a) no solution bifurcates and 
(b) no new
 solution appears. 
 This can take place either  as a global property of all TAP states, or
 as a property 
of a restricted subset of dominant states, selected according to
 their weights.
Under these  hypotheses this means that we may write
\begin{equation}
\left[ g(m(h)) \right]_{\rm sol} =\frac{1}{{\cal N}_s}
\sum_{\alpha}g(m^{\alpha}_i(h))
\label{ipoBRST}
\end{equation}

As a consequence, for any $n$, the following relation holds
\begin{equation}
{\partial^n \left[ g(m(h)) \right]_{\rm sol} \over \partial h^n}
=\frac{1}{{\cal N}_s}\sum_{\alpha}{\partial^n g(m^{\alpha}_i(h)) 
\over \partial h^n}
\end{equation}
We start considering $g(m)=m_k e^{\beta u F(m)}$, which is 
 the average
 of the magnetization with the weight $e^{\beta u F}$, used to count
 solutions of a given energy. In general we have:
\begin{equation}
{\partial \left[ m_l e^{\beta u F(m)} \right]_{\rm sol} 
\over \partial h_k}
=-\left< x_k m_l  \right>
\end{equation}
where the average on the R.H.S. is computed with respect to the
 action $S(x,m, \overline{\psi},\psi )$ [Eq. (\ref{actionpsi})].
However,
under the above hypothesis, this must be equal to the average 
on each solution of the derivative of  $m_k e^{\beta u F(m)}$ with
 respect to $h_k$, which is 
\begin{equation}
\sum_{\alpha}\left( { \partial m_l^\alpha \over \partial h_k}
e^{\beta u F(m^\alpha)} 
+m_l \beta u {\partial F(m^\alpha) \over \partial m_j^\alpha}
{\partial m_j^\alpha \over \partial h_k}\right)
\end{equation}
We can drop the second term since, by definition, it is zero on the 
TAP-solutions,  and we are left with 
\begin{equation}
\frac{1}{{\cal N}_s}\sum_{\alpha}\chi_{kl}e^{\beta u F(m^\alpha)}
=\frac{1}{{\cal N}_s}\sum_{\alpha}
\left({\partial^2 F\over \partial m^\alpha \partial m^\alpha}\right)^{-1}_{kl}
e^{\beta u F(m^\alpha)}=\left< \overline{\psi}_l \psi_k \right> 
\end{equation}
Thus we have
\begin{equation}
-\left< x_k m_l \right>=\left< \overline{\psi}_l \psi_k \right> 
\end{equation}
i.e., for $k=l$, the first Ward identity for BRST-susy systems, Eq. (\ref{W1}).
To obtain the second Ward identity, Eq. (\ref{W2}), 
we must apply the same procedure to the second derivative of 
the quantity $g(m)=e^{\beta u F(m)}$, proportional to the complexity.
From Eq. (\ref{actionpsi})
we have:
\begin{equation}
\frac{\partial^2 \left[e^{\beta u F(m)}\right]_{\rm sol}} 
{\partial h_i \partial h_j}=\beta \left< x_i x_j \right>
\end{equation}

Under the hypothesis of Eq. (\ref{ipoBRST}),
 this must be equal to the following quantity:
\begin{equation}
\frac{1}{{\cal N}_s}
\sum_{\alpha}{\partial^2 e^{\beta u F(m^\alpha)} 
\over \partial h_i \partial h_j}
=\frac{1}{{\cal N}_s}
\sum_{\alpha}\left(\beta u{\partial^2 F(m^\alpha)  \over \partial h_i \partial h_j}
+\beta^2 u^2{\partial F\over \partial m_l}{\partial m_l \over \partial h_i}
{\partial F\over \partial m_s}{\partial m_s \over \partial h_j}\right)e^{\beta u F(m^\alpha)}
\end{equation}
where the second term  is zero over any TAP solution and we are left with 
\begin{equation}
\sum_{\alpha}\beta u{\partial^2 F(m^\alpha)  \over \partial h_i \partial h_j}
e^{\beta u F(m^\alpha)}=\beta u \left< \overline{\psi}_i \psi_j \right> 
\end{equation}
Thus we obtain
\begin{equation}
\left< x_i x_j \right>
=u \left< \overline{\psi}_i \psi_j \right> 
\end{equation}
The previous BRST relations can be obtained under weaker
 conditions than those considered initially, in particular
the above derivation still holds if we allow the onset
of  new solutions, provided that they appear {\em  only
 outside} the ensemble 
of dominant solutions, e.g. at the threshold. 
Indeed, 
 in presence of a weight $e^{\beta u F(m)}$, only solutions 
of a certain free energy count, thus supporting our assumption that,
in the dominion of interest, the 
total number of states, ${\cal N}_s$, does not change.
Instead, the condition of no-bifurcation seems unavoidable to obtain 
the BRST relations by means of this argument.

\subsection{Identification of BRST solution with static solution. 
Link with Replica computation.}

\label{s:1rsb}

In this section we very shortly recall the connection between the static 
solution at the one-step RSB approximation  and the BRST-susy 
solution of $\Sigma_a$ \cite{CGPM}.
We compare results at given values of the breaking parameter $m$
in replica formalism and at given values of the variable $u$, 
Legendre conjugated of 
$f$, in the framework of the annealed computation of the complexity.

The total replica free energy at one-step RSB  is:
\BEQ
f^{\rm rep}=-\frac{\beta}{4}\left(1-2 q_1\right)
-\frac{\beta}{4}\left[
(1-m)q_1^2+m~q_0^2\right]-\frac{1}{\beta~m}
\int{\cal{D}}z_0\log\int{\cal{D}}z_1~p_1(m,z_0,z_1)
\label{1:Phi}
\EEQ
with
\BEQ
p_1(m,z_0,z_1)\equiv \left[2\cosh\left(\beta z_0\sqrt{q_0}
+\beta z_1\sqrt{q_1-q_0}\right)\right]^m
\label{1:p1}
\EEQ
Self consistency equations take the form 
\BEA
&&q_0=\int{\cal{D}}z_0 \left<\tanh\left(\beta z_0\sqrt{q_0}
+\beta z_1\sqrt{q_1-q_0}\right)\right>^2
\label{1:q0}
\\
&&q_1=\int{\cal{D}}z_0 \left<\tanh^2\left(\beta z_0\sqrt{q_0}
+\beta z_1\sqrt{q_1-q_0}\right)\right>
\label{1:q1}
\\
&&
\frac{\beta}{4} \left(q_1^2-q_0^2\right)
+\frac{1}{\beta m^2}\int{\cal D}z_0\left[
\log\int{\cal{D}} z_1~ p_1(m,z_0,z_1)
-\left<\log p_1(m,z_0,z_1)\right>
\right]=0
\label{1:m}
\EEA
with
\BEQ
\left<\left(\ldots\right)\right>\equiv
\frac{\int{\cal{D}}z_1 ~\left(\ldots\right)~p_1(m,z_0,z_1)}
{\int{\cal{D}}z_1 ~p_1(m,z_0,z_1)}
\label{1:ave}
\EEQ

Leaving $m$ as a free parameter (thus ignoring 
Eq. (\ref{1:m})),   at a given temperature we can define the complexity 
of the system as 
the  Legendre transform of 
$\beta m~f^{\rm rep}$:
\BEQ
\Sigma_1(f) = \max_m \left[\beta m f - \beta m~f^{\rm rep}(m)\right]
\label{LegTra}
\EEQ
with conjugated variables $\beta m$ and $f$,
\BEA
&& f =\frac{\p~ m f^{\rm rep}}{\p m}
\label{f_con}
\\
&& \beta m = \frac{\p \Sigma_1}{\p f}
\label{m_con}
\EEA
Eqs. (\ref{f_con}) and  (\ref{m_con})  yield the relation between $f$ and $m$.
Introducing Eq. (\ref{f_con}) in the Legendre transformation of 
$\beta m~f^{\rm rep}$ [Eq. (\ref{LegTra})] one can obtain 
the following relation:
\BEQ
\Sigma_1(f) = \left.\beta m^2 \frac{\p f^{\rm rep}}{\p m}\right|_{m(f)}
=\left\{\frac{\beta^2}{4} m^2 \left(q_1^2-q_0^2\right)
+\int{\cal D}z_0\left[
\log\int{\cal{D}} z_1~ p_1(m,z_0,z_1)
-\left<\log p_1(m,z_0,z_1)\right>
\right]
\right\}_{m(f)}
\EEQ

In zero external magnetic field  $q_0=0$. In this particular case 
$\Sigma_1$ represents the  entropy of {\em hidden states} that we 
would get with the method of the $m$ coupled real replicas \cite{MPRL95}
and 
a formal connection can be established between the complexity
of the BRST-susy saddle point of the annealed computation and 
the one step RSB replica free energy.
This was shown in Ref. \cite{CGPM} 
[Eqs. (\ref{rho_CGPM})-(\ref{fen_CGPM})] but the formal connection
also holds  in the original notation of Ref. \cite{BMan}.

Indeed, the expression of the annealed
 complexity obtained simplifying $\Delta$ and
$\lambda$ in the logarithm of Eq. (\ref{rho_BM})
by means of the BRST relations Eqs. (\ref{BRST1}), (\ref{BRST2})
 comes out to be
\BEQ
\Sigma_a(f; u, q)=\log \rho_s(f; u, q, \Delta(u,q), \lambda(u,q))
=\beta u\left\{
-\frac{\beta}{4}\left[1- 2q+(1+u) q^2\right]
+\frac{1}{\beta u}\log\int{\cal{D}} z\left[2\cosh \beta z\sqrt{q}\right]^{-u}
-f
\right\}
\label{BRST_comp}
\EEQ

It can be easily seen that, putting $q_0=0$ in the replica free
energy, Eq. (\ref{1:Phi}), this is bounded to Eq. (\ref{BRST_comp})
by

\BEQ
\Sigma_a(f; u, q) = \beta   u \left[f^{\rm rep}(q,-u) - f\right]
\label{1:identS}
 \EEQ
provided that one makes the identification
\BEQ 
m = -u     \hspace*{3 cm}   q_1 = q 
\label{ident}
\EEQ

Furthermore, the saddle points of such a function 
$\Sigma_a$  coincide with those 
of $f^{\rm rep}$,  if
the correspondence $f^{\rm rep}(q, -u=m^\star)= f$
holds 
 at the static value of the breaking
parameter $m$, i.e. $m^\star$ satisfies 
Eq. (\ref{1:m}).

The identification in Eq. (\ref{ident}) is evident 
for the overlap: $q_1$ is the self-overlap of states
in the replica 1RSB framework
and $q$ is the self-overlap  of TAP-states, i.e. they are just different 
representations of the same thermodynamic observable.
 For what concerns the 
connection between the breaking parameter $m$ and the variable $u$
conjugated with the  TAP free energy, we can observe that
the derivative with respect to $f$ of Eq. (\ref{BRST_comp})
gives
\BEQ
\frac{\p \Sigma_a}{\p f} = -\beta u
\EEQ
Simply comparing this with Eq. (\ref{m_con}) one can identify $-\beta u$ as
 $\beta m$, the slope of the complexity as a function of the free energy.
As a matter of fact we are just saying that Eq. (\ref{BRST_comp})
can be seen as the Legendre transform of ($-\beta u~ f^{rep}$) with
conjugated variables $-\beta u$ and $f$.

Notice that, as a consequence of the link 
shown in Eqs. (\ref{1:identS}), (\ref{ident}),
the average defined in Eq. (\ref{ave}) coincides with the one defined in Eq.
(\ref{1:ave}). Therefore we used the same symbol.

The one-step RSB static solution
 is not thermodynamically stable, as it can
be shown computing the Hessian of the replica thermodynamic potential with
respect to the order parameters variations around their self-consistently 
derived values, given by Eqs.
(\ref{1:q0}), (\ref{1:q1}). 
Indeed, the eigenvalue associated with the overlap-overlap
fluctuations, i.e. the replicon, is
\BEQ
\Lambda_0= 
1-\beta^2  \left<
\cosh^{-4}(\beta  \sqrt{q_1} z_1)\right>
\EEQ
Using the identity  $\cosh^{-4}=(1-\tanh^2)^2$, the stability condition 
$\Lambda_0 > 0$ can be written as
\BEQ
T^2 > 1- 2 q_1 +\left<\tanh^4(\beta  \sqrt{q_1} z_1)\right>
\EEQ
This is the analogous in the replica formalism of the Plefka's 
criterion for the physical relevance
 of TAP solutions \cite{Ple0} (see Sec. \ref{ss:ple})

As we can see from Fig. \ref{f:4}, the replicon of the one-step RSB solution
at zero magnetic field, as a function of $u = -m$, is always negative
in the dominion where the complexity
 is positive, even on the 'static' point $f_0
\equiv f(m^\star)$.
From the same figure, though, we observe that the equivalent quantity
computed over the BM saddle point of the annealed complexity
is such that Plefka's criterion is always satisfied. This is 
a necessary condition
 supporting the possibility that the BM annealed complexity could
represent the number of states of the SK model, yet not a sufficient one,
as we will see in Sec. \ref{s:stab_anneal}.

  \subsection{The Two-group Ansatz breaks  the BRST-susy}
  \label{s:2G}
 In Ref. \cite{BM2g} Bray and Moore introduced the 
{\em two-group Ansatz} in order to solve the instability problem
of the replica symmetric solution of the SK model.
In Ref. \cite{MPRL95} Monasson showed how the formalism of Legendre
transforms can be applied to mean-field disordered models through
the {\em pinning of real replicas} in a configuration space
extended to $m$ copies of the system.

Parisi and Potters \cite{PP} explained how the BM action can be obtained 
through the method of Monasson provided that the symmetry between
real replicas is broken  according to a generalized two-group Ansatz. 
Being  $n$ the number of replicas introduced to compute the quenched average
(following the standard scheme \cite{SKPRL75,MPV}) 
and $m$ the number of real copies,
they analyzed 
\BEQ
\lim_{n\to 0}\frac{1}{n}\log{\overline{ Z^{mn}}}
=\ext_f\left[{\overline{\log \rho_s(f)}}- m\beta N f\right]
\label{logZmn}
\EEQ
where $Z^{mn}$ is the partition function of $n\times m$ copies of the system.
In terms of the replicated matrix-parameter $\hat Q$ its average is
\BEQ
{\overline{ Z^{mn}}}=
\ext_{\hat Q} \exp\left\{N\left[\frac{\beta^2}{4}
\left(mn-\Tr {\hat Q}^2\right) + \log\sum_{\{\sigma_a^c\}}
\exp\left(\frac{\beta^2}{2}\sum_{ab,cd}{\hat Q_{ab}^{cd}}
\sigma_a^c\sigma_b^d\right)
\right]\right\}
\label{Zmn}
\EEQ
where the indexes $a, b = 1,\ldots, n$
while $c, d= 1,\ldots, m$. 
The four index matrix ${\hat Q}_{ab}^{cd}$ 
can be expressed as the composition of $n^2$ sub-matrices $\mathbf{Q}_{ab}$ 
of dimension
$m\times m$ of the form:
\BEQ
\mathbf{Q}_{ab}=\overbrace{\left( \begin{array}{l}Q^+_{ab}  \\
				         Q_{ab}  
			   \end{array}
			   \right.}^{y}
       \overbrace{\left. \begin{array}{r} Q_{ab}  \\
			                  Q^-_{ab}
			   \end{array}
			   \right)}^{m-y}
\label{Q2g}
\EEQ
The matrices $Q_{ab}^{\pm}$ are further parameterized as 
\BEQ
Q_{ab}^{\pm}=Q_{ab}\pm \frac{A_{ab}}{y}+\frac{C_{ab}}{2 y^2}
\EEQ
Furthermore is $\hat Q_{aa}^{cc}\equiv 0$.
In Ref. \cite{PP} the last term was $B_{ab}/y^2$. We write
$C_{ab}= 2 B_{ab}$ both in order to avoid confusion
with the parameter $B$ in the $\Sigma_a$ expression of Sec. \ref{s:annealed}
and to obtain more symmetric expressions in following works.

Eq. (\ref{logZmn}) is, then,  computed making use of the above Ansatz, getting 
a complexity that can be formally connected to the one of 
Bray, Moore and Young, Ref. \cite{BMY} through a given change of variables.
We will  analyze it more explicitly  elsewhere \cite{CLPR},
 where we will use
such transformation.

For the time being we are mostly interested in the annealed case,
that can be obtained putting in matrix (\ref{Q2g}) all $Q_{ab}$ with 
$a\neq b$ equal to zero. The Ansatz reduces, then, to consider 
$n$ diagonal blocks of $m\times m$ matrices (\ref{Q2g}), for $b=a$, 
where the elements are built with: $Q_{aa}=Q$, $A_{aa}=A$, $C_{aa}=C$.
In this case, the change of variables we need to connect the BM
formulation of the annealed complexity to the Parisi-Potters (PP)  one, are
\footnote{Notice that also in Ref.\cite{BMque,BMY}
there is a $Q$. In Ref. \cite{BMY} this is related to $q$ (TAP self-overlap)
as $Q=\beta^2 q$.
The parameters $Q$ in \cite{BMque,BMY} ($Q_{\rm BM}$) and in \cite{PP} 
($Q_{\rm PP}$)
are proportional to each other but different:
$Q_{\rm BM}=\beta^2 Q_{\rm PP}$.}:
\BEA
\mbox{BM} && \mbox{PP}
\nn
\\
q&=&Q
\label{PP_BMq}
\\
\Delta &=&\left(A+\frac{m}{2}Q\right)
\label{PP_BMD}
\\
\lambda &=&\frac{\beta^2}{2}\left(C+ m A+\frac{m^2}{4} Q\right)
\label{PP_BMl}
\\
u &=&-m
\nn
\EEA

Writing the equations with the substitutions 
(\ref{PP_BMq})-(\ref{PP_BMl}), allows for an immediate connection
between breaking the matrices structure into two groups
 and  breaking the BRST-susy.
Indeed, Eqs. (\ref{PP_BMD}) and (\ref{PP_BMl}) transform into the
BRST relations (\ref{BRST1}) and (\ref{BRST2}) if
we set $A=C=0$, i.e. if we do not break the matrix structure at all 
({\em unbroken two-group}).
On the contrary, setting values of A and C different from zero amounts to
break the BRST symmetry and leads to values of $q$, $\Delta$ and $\lambda$
independent among each other, i.e. to the BM solution.

\section{Properties of the Annealed Complexities}
  \label{s:stab_anneal}
 In the previous part of the paper, we have studied
 the so called BM complexity \cite{BMan} (the one breaking the supersymmetry)
and the BRST-symmetric complexity \cite{CGPM} as two different
solutions of the same saddle point
equations, derived from the same BRST-susy functional
(see Sec. \ref{ss:inva}).

In this section we will discuss several issues connected to the problem of 
selecting a proper, physically meaningful, complexity and to see whether
any of the two candidates fulfill the minimal requirements.

We start noticing that not all TAP solutions can be associated to stable 
thermodynamic states.
They have first to satisfy the Plefka's criterion (see below),
which guarantees the right expression for the 
linear susceptibility of the states under consideration,
and, then, to be minima of the TAP free energy 
functional.

Operationally speaking, 
what happens
 is  that, even if on physical
 grounds we are willing
to count only the minima of the TAP free energy, in practice,
we also count other kind of solutions of the TAP equations. 
As a consequence, it may happen that
 we have a mathematically correct computation which, however, has no physical 
meaning at all or a rather obscure one.   
In  subsection \ref{ss:ple} we will see that, at the present stage,
we have no way to
 state that any of the two solutions of the 
complexity saddle point equations 
correspond to counting only minima of the TAP free 
energy. 
On the other hand, though it is just a necessary condition,
 a partial discrimination can be done on the basis of Plefka's
criterion, which 
  is satisfied by the BRST-susy-breaking
 solution but not by the BRST-susy solution.

Since the BRST solution can be linked to the saddle point  solution 
obtained in a 1RSB
 replica computation in zero magnetic field, it is natural to check
 its stability according to the
 usual criteria in that framework, 
e.g. the positivity of the replicon eigenvalue; 
however, as we anticipated in Sec. \ref{s:1rsb},
this condition is formally identical to the Plefka criterion,
which is, therefore 
 violated by the BRST-susy solution.  This violation leads to a
 mathematical inconsistency, as we show
in Sec. \ref{ss:ple}.

The  BRST supersymmetry is a property of the action and, therefore,
 its violation
 poses a  consistency problem to the BM saddle
 point, too, due to the lack of control on what we are counting. 
The only possibility for the BRST-susy-breaking
 saddle point to be valid is to guess that
 it counts only the minima, as it is claimed in Ref. 
\cite{commentBM}.
 It would describe only a physically stable
portion of the set of TAP solution and we
 would not expect that it satisfies global properties valid on the whole set,
 such 
as the Ward identities derived from the BRST susy [see Eqs. (\ref{W1})-
(\ref{W2})].
However, this assumption is hard  to justify, as we will discuss in 
the subsections \ref{ss:spectrum}-\ref{ss:det}, 
recalling the properties of the  Hessian of the 
TAP free energy as derived in Ref. \cite{Ple1}.

The  effects of neglecting the sign of the determinant of the TAP solutions,
 which is at the origin of the BRST susy, are particularly dramatic when we
 set $u=0$.
In this case we are summing over each extremum of the TAP functional with a
 weight given by the sign of its determinant. 
The Morse theorem tells us that 
this quantity is a topological invariant and it is equal to $1$ in
 this specific case. While the BRST-susy solution does satisfy the theorem,
the BRST-susy-breaking
 saddle point yields a quantity of order $e^{N \Sigma_{\rm BM}} \gg 1$, if 
$\Sigma_{\rm BM} >0$.

A first way to recover mathematical consistency in the $u=0$ case is to guess
 that BRST susy is restored, e.g.  because  the prefactor of the 
exponential  is  $e^{-N \Sigma_{\rm BM}}$.
In this case the expansion of the prefactor in powers of $1/N$ will be zero 
at all orders (such as the expansion of $e^{-1/x}$ in powers of $x$). 
This series was considered by Kurchan in Ref. \cite{K91} were it is shown that 
its coefficients are indeed all null.
This proves that the BM saddle point is possibly mathematically  consistent 
but rises another difficulty.
We already stressed that at the order $N$ it is not possible to 
prove whether the BM complexity is counting only minima or not.
When computing the corrections without modulus of the determinant 
of the Hessian
one seems to have a vanishing prefactor, though. 
Indeed, at this stage, in order to accept the BM prediction, we
 must, then, assume that the effect of taking the modulus 
of the determinant into account is to
 change the value of the prefactor from  exponentially small  to  finite,
 a completely unjustified assumption  at the present state.

 

\subsection{Plefka's criterion}
\label{ss:ple}
In this subsection we recall the results of Ref. \cite{Ple1} on  
the properties of the Hessian of the TAP free energy on a generic 
configuration $\{ m_i \}$, which is also the inverse of the susceptibility 
matrix:
\BEQ
\chi_{ij}^{-1}=\frac{\p^2 F_{\rm tap} (m)}{\p m_i \p m_j}
\EEQ
The problem is studied considering  the resolvent of the Hessian. As a
 byproduct of the computation one obtains the magnetic susceptibility of a
 solution. On physical grounds this must be equal to $\beta (1-q)$, 
however it turns out that not all TAP configurations satisfy this relation.
 Instead the condition for a TAP solution $\{ m_i \}$ to yield the right 
physical susceptibility is  
\BEA
&& x_{\rm P}\equiv 1-\beta^2\frac{1}{N} \sum_i (1-m_i^2)^2 
\geq 0
\label{plefkac}
 \EEA
The TAP solutions which do not satisfy the previous relation turn out 
to have the 
following non-physical susceptibility:
\BEQ
\chi_l=\beta (1 - q)+\frac{x_{\rm P}}{p}
\label{malachi}
\EEQ
where $p$ is defined as
\BEQ
 p\equiv \beta^3\frac{1}{N}\sum_i(1-m^2)^3
\EEQ
Therefore, after having computed the complexity, we must check that the Plefka
 relation is verified to be sure that we are counting physical solutions.

For $N\to \infty$, the site-average becomes, 
in the notation of the annealed computation of Sec. \ref{s:SPan}, the
 average of Eq. (\ref{plefkac})  over the action ${\cal{L}}$ 
[Eq. (\ref{action})] and the above 
inequality reads:
\BEQ
1-\beta^2\left(1-2 q +\left<m^4\right>\right) \geq 0
\label{plefkac2} 
\EEQ
As we can see from Fig. \ref{f:4} such a condition is satisfied
by the saddle point breaking the BRST-susy, but is violated by the
BRST-invariant 
one.
In the replica language Eq. (\ref{plefkac2}) is the replicon.
The identification of the measures over which the average is performed
is given by Eqs. (\ref{BRST_comp})-(\ref{ident}).

As we  show in  Appendix B, the computation of the determinant of the
 Hessian, which is a crucial step in the derivation of the action, is deeply
 connected to the Plefka computation of the resolvent \cite{Ple1}. 
As a consequence, one 
sees that, in order to set the parameter $B$ of the determinant 
equal to zero,
as it was chosen in  both solutions, one 
must check that  Plefka criterion is satisfied. 
Therefore,  the fact that the BRST-susy  solution violates it, 
  turns out to be not only a problem of physical meaning, but also  a problem
 both  for the  replica computation and for the  mathematical consistency of
 the solution.
 
We conclude stressing that  although the annealed computation is well 
defined on mathematical grounds, what really matters for physics  is the
 quenched computation where this problem can  possibly be cured considering a 
full-RSB BRST-susy solution. 
As we will see, this certainly happens at the 
lower band-edge of the quenched complexity (the free energy value at which 
the complexity vanishes) which, 
as expected, coincides with the equilibrium free 
energy  given by the Parisi solution and it is BRST symmetric.

\subsection{The spectrum of the Hessian of $F_{\rm tap}(\{m\})$}
\label{ss:spectrum}
The spectrum of the eigenvalues of the inverse susceptibility matrix
${\mathbf{\chi}}^{-1}(m)
$ on a generic configuration $\{ m_i\}$ is determined in Ref.
\cite{Ple1} for small eigenvalues and in the region 
of $\{m\}$ configurations such that
$x_{\rm P}\simeq 0$.
The eigenvalues distribution is written, at the leading order in $N$, as
\footnote{In Ref. \cite{Ple1} the factor 1/4 in front of 
the second term in the square root was missing.}:
\BEQ
\rho(\lambda)=\frac{1}{\pi\sqrt{p}}\sqrt{\lambda
-\frac{x_{\rm P}^2}{4p}},
\hspace*{1 cm} |x_{\rm P}| \ll 1, \hspace*{1 cm} \lambda \ll 1.
\EEQ
The minimum of the spectrum is therefore positive, irrespective 
of the value of $x_{\rm P}$, implying 
that the TAP free energy is semi-convex everywhere at the
extensive level.
In the same paper this result was proven rigorously also 
for generic  configurations with a finite $x_{\rm P}$, either positive or 
negative. 
We remark that this result, 
however, {\em does not exclude the existence of solutions
 with a sub-extensive number of negative eigenvalues}.
Actually, Morse theorem implies that these solutions do exist and,
indeed, there must be  an equal number of solutions with positive 
and negative determinant. Plefka's result just states that  we cannot 
distinguish them at an extensive level.


\subsection{The determinant of the Hessian}
\label{ss:det}

As we have previously discussed,
the fact that no negative eigenvalue
exists in an extensive quantity does not mean that saddles and maxima 
do not exist, which is indeed impossible on topological grounds.
Therefore, the condition 
$x_{\rm P}\geq 0$ does not bypass the problem
of the sign of the determinant of $\mathbf{\chi}^{-1}$.

In appendix A the determinant is computed making use of Grassmann
variables and using the saddle point method.
If one evaluates the saddle point over $B$ before evaluating the integrals 
over the $m_i$, it turns out that there are two solutions $B=0$ and $B>0$ 
($B$ is the parameter originally used 
in Ref. \cite{BMan}
\footnote{There, the quenched average determinant of the Hessian was computed 
 as
\BEQ
{\overline{\det{\mathbf \chi}^{-1}}}
= \ext_B
\exp\left\{N\left[\frac{B^2}{2\beta^2}- B(1-q)+\frac{\beta^2}{2}(1-q)^2
+\frac{1}{N}\sum_i\log\left(\frac{1}{1-m_i^2} +B\right)
\right]
\right\}
\EEQ
where 
the initially integer
positive number of replicas was sent to $-2$ at the end of the computation.}). 
If $x_{\rm P}\geq 0$ the solution $B=0$ satisfies 
the condition for the saddle point of being a maximum of the integrand on 
the integration path and is the correct one, while if $x_{\rm P}<0$ the
 correct solution is the one with $B>0$.

Both the BRST-symmetric and the non-BRST-symmetric solutions
 assume that $B=0$, consequently the  expression of the determinant of the Hessian of the TAP 
free energy becomes
\BEQ
{\overline{\det{\mathbf {\chi}}^{-1}}} = \prod_i\left(\frac{1}{\beta}
\frac{1}{1-m_i^2}\right)
\exp\left\{N\frac{\beta^2}{2}(1-q)^2
\right\}
\EEQ
 Since the prefactor is positive for any $\{m_i\}$ configuration,
this choice of $B$ {\em would provide} an {\em a posteriori}
 justification for neglecting the absolute value of the determinant
{\em if the calculation were exact} at all orders in $N$.
Anyway, we cannot neglect the fact that all computations
are performed at the leading term in $N$, as $N\to \infty$,
thus ignoring all sub-extensive contributions.

As a matter of fact, the sign also depends on neglected prefactors:
 \BEQ
{\overline{\det{\mathbf \chi}^{-1}}}
= \alpha(N) \left.{\overline{\det{\mathbf \chi}^{-1}}}\right|_{\rm sp}
\EEQ
where $\alpha(N)$ cannot be  determined at the leading order in $N$, but 
needs to be obtained from corrections of $O(1)$:
\BEQ
\frac{1}{N}\log  {\overline{\det{\mathbf \chi}^{-1}}}=
\frac{1}{N}\log  \left.{\overline{\det{\mathbf \chi}^{-1}}}\right|_{\rm sp}
+\frac{1}{N}\log \alpha(N)
\EEQ

While the magnitude of the determinant is not qualitatively changed
by $O(1)$ corrections (provided they are not zero, of course), its sign can
well be determined by eigenvalues that are present in a sub-extensive
quantity.

\subsection{The Bray-Moore-Young FRSB quenched complexity
 is BRST-invariant.}
\label{ss:bmy}
Both the BRST-susy and the BRST-susy-breaking solution give a
  lower band-edge (the free energy value at which the complexity vanishes)
 different from the equilibrium free energy of the SK model. This is not 
surprising since we are performing annealed averages, while we expect that
 the physically correct computation is quenched.  

In order to cure this deficiency it is rather obvious that one has to 
go on computing the quenched complexity in the Full RSB scheme, 
for which the SK model is known to be thermodynamically stable. 
The quenched case is formally  discussed in Ref. \cite{Alessia}, 
and one instance of FRSB quenched complexity is presented in Ref. 
\cite{CLPR}.   
The first step in this direction, however, was  
performed by Bray, Moore and Young in Ref. \cite{BMY}
where the solution there analyzed had a lower band-edge coinciding with
the FRSB static one. 
Now, we notice that the assumptions under which they look for such 
a solution, 
are exactly the BRST relations (\ref{BRST1}), (\ref{BRST2}) generalized to
the function order parameters one has to introduce in the quenched FRSB 
scheme 
of computation.
Since we are using their very same notation we can  directly rewrite here
Eq. (19) of Ref. \cite{BMY}, which we break into two lines:
\BEQ
\Delta = -\rho(1)
\hspace*{1 cm}
\lambda=\frac{\beta^2}{8}u^2 q_{EA}
\label{BMY1}
\EEQ
for the diagonal part (the one surviving in the annealed case),
and
\BEQ 
\rho(x)=\frac{\beta^2}{2}u~ q(x)
\hspace*{1 cm}
\rho^{\star}(x)=\frac{\beta^2}{4}u^2 q(x)\hspace*{1 cm} \forall x\in[0,1]
\label{BMY2}
\EEQ 
 for the off-diagonal elements. The further assumption was, then, made
\BEQ
q(1)=q_{EA}
\label{BMY3}
\EEQ

The relations expressed in Eq. (\ref{BMY1}) [inserting  assumption
 (\ref{BMY3})]
are exactly our Eqs. (\ref{BRST1}), (\ref{BRST2}).
The same holds 
 for the off-diagonal terms if we recognize that, in Eq. (\ref{BMY2}),
the off-equilibrium analogue of $\Delta$ is $-\rho(x)$ and
the analogue  of $\lambda$ is $\rho^\star(x)/2$.
What was found is, then, the quenched improvement of the 
annealed BRST-susy solution and not
the quenched analogue of the BM annealed
 solution of Ref. \cite{BMan}, for which
the above mentioned relations do not hold.
Moreover, the choice of identifying the elements $q(1)$ 
of the diagonal block of the FRSB
matrix   with the elements on the diagonal, $q_{EA}$, leads 
precisely to the self-consistent, stable, Parisi solution of the SK 
model \cite{P80}. 
In such a case, though, no parameter is left free to vary
and, therefore,
no analysis over the number of states at given $f$ can be performed,
i.e. no complexity can be built.

It is the subject of another paper
\cite{CLPR} to go beyond this point and look for a generalized
solution that allows a 'quenched probe' in 
a free parameter (the state free energy $f$ or the generalization 
of the $m$ parameter in the Legendre transform approach), yet recovering
the right equilibrium value as lower band-edge
of the complexity.



\section{Conclusions}

In this paper we have shown that the BM action \cite{BMan} 
and the  action considered by Cavagna {\em et al.} \cite{CGPM} are completely equivalent 
and one can move from one to the other through a simple change of variables.
This is also true at the microscopic level thus implying that the BM action
 too 
is BRST-susy invariant.
In particular, this equivalence implies that each solution of the BM
 saddle point 
equations is a solution of the equations of Ref. \cite{CGPM} 
as well and {\it vice-versa}.
As a consequence we are left with the problem of selecting the ``right'' 
solution.
 
In order to determine which one, if any, of the two proposals was the one 
actually representing the multiplicity of stable and metastable states
of the Sherrington-Kirkpatrick mean-field spin-glass model
we have been critically reviewing the properties of both, sometimes solving
apparent incongruencies, other times  pointing out substantial inadequacies.

We first summarize the case of the  BM BRST-susy breaking annealed complexity.
The BM saddle point is not BRST-susy at any value of $u$, 
while we have shown in Sec. \ref{s:annealed}
that the BM action satisfies such a supersymmetry,
 although this is differently expressed in the BM notation with respect to the
one of Ref. \cite{CGPM}. 
As already noticed  by Kurchan in Ref. \cite{K91}, we have to be careful and 
we need some justification before adopting it, if we want to preserve  
mathematical 
consistency. 
In particular, one may show that the BRST-susy
 is restored considering subextensive 
corrections to the saddle point  and notice
that they amount to a prefactor of $\exp N \Sigma_{\rm BM}$.
In the case $u=0$, it has been shown by Kurchan by means of a series 
expansion in power 
of $1/N$ that   the prefactor of the 
 non-BRST saddle point is zero at all orders of the  expansion. 
This could imply
a zero prefactor or it could leave
  the way open to a non zero but exponentially small prefactor. Both outcomes, 
 however, strongly change the BM prediction. 
Therefore, to save 
this prediction, one should show that the effect of keeping the modulus of the
 determinant amounts to change the prefactor from an exponentially small value
 to a finite one. A step in this direction could be possibly done
generalizing 
the technique  of Ref. \cite{CGG2} for one dimensional random 
systems.

Very recently, in Ref. \cite{commentBM},
 it has been claimed that the BM saddle point counts only minima. 
This would imply that  
the complexity does not involve a sum over all 
solutions and, therefore,  is not constrained to satisfy global relations, 
like those imposed by
 BRST-susy or by the Morse theorem.
However, there is no prove at all that the BM saddle point counts only minima,
 since
Plefka's analysis
 of the Hessian shows that all solutions of the TAP equations have 
strictly positive eigenvalues only at the leading order in $N$. 
In order to get information about the sign of the determinant one should
be able to select the minima among all states satisfying
Plefka's criterion. This can be explicitly seen in the spherical $p$-spin
models where these solutions can be classified as minima and saddles,
differing only for one negative eigenvalue.
The Plefka criterion, $x_{\rm P} \equiv 1-\beta^2(1- 2 q + 1/N\sum_i m_i^4) >0$
 is, thus,
  not related to the fact that a 
given solution is a minimum or a maximum or a saddle but rather it guarantees 
that it yields the correct
 susceptibility.

The lower band-edge of
 the quenched complexity, 
computed with the Parisi Ansatz by Bray, Moore and Young \cite{BMY}, 
gives the correct equilibrium free energy  but such solution
 is not a modification of the  BRST-susy breaking annealed
 solution
of Bray and Moore \cite{BMan}.
On the contrary, it turns out to be BRST symmetric, as we have shown in Sec.
 \ref{ss:bmy}.
This also means that, up to now, 
no quenched extension of the BM annealed saddle point has been taken into
 account.
We will show elsewhere what are the minimal assumptions to break 
the BRST-susy  in the quenched case \cite{CLPR}.

Furthermore, the non-BRST solution is not in agreement with the numerical
 results recently obtained by Plefka through some modified TAP equations 
\cite{Ple2}. 
By  means of its method he  obtains all the  minima of the standard TAP
 free energy satisfying $x_{\rm P}\geq 0$. 
The special set of solutions he
 collects,
 i.e. the minima with $x_{\rm P}\geq 0$, is precisely the set that is supposed
to be counted by the BM complexity. Now, the BM solution predicts that
 the great majority  of  solutions has a value of $x_p$, e.g. at temperature 
$T=.5$ given by $x_p= 0.132408 > 0$ ($\Sigma_{\rm BM}=0.002775$) and at 
$T=.2$, by $x_p= 0.5952975 > 0$ ($\Sigma_{\rm BM}=0.05219$).
Plefka, however, on the basis of  its
 numerical data, hints that all the minima of the TAP functional in the
 thermodynamic limit have a  zero $x_p$ as $N\to \infty$. 
Moreover, the number
of solutions found by Plefka varying $N$ are, e.g. at $T=0.2$,
 less then $\exp N\Sigma_{\rm BM}$
and the discrepancy increases with increasing $N$. This is also true
for what concerns the range of free energy values over which the complexity
is non zero. This finding is also in agreement with 
what has been found at the FRSB level of computation,as will be discussed in
\cite{CLPR}.

Notice also that the BM 
result for $x_{\rm P}$ cannot be changed considering the quenched average 
instead of the annealed one. In Ref. \cite{BMan} also the replica symmetric 
quenched
complexity was  considered  and there the authors showed that  the annealed
 and the quenched  
non-BRST saddle points coincide
 at $u>u_c$ ($u_c<0$) and, in particular, at $u=0$ 
to which  the highest number of solutions
would correspond if the modulus of the determinant of the Hessian 
was taken into account.

It is worth mentioning, anyway, that the zero temperature limit of the BM
total complexity coincides with the  computation of 
the number of solutions of the zero temperature limit of the TAP  equations,
$m_i =\mbox{sign}(\sum_{ij}J_{ij}m_j)$, where no reaction term is  present
 \cite{BMan,TanaEd,DDGGO}.
If the coincidence of the exactly zero temperature behavior with the
$T\to 0$ behavior would be a necessary condition, this would be a strong hint 
that the BM saddle point provides, indeed, the right complexity.
However, we notice that
this does not exclude the existence of other solutions
displaying a $T \to 0$ limit of the complexity different from
the value directly computed at $T=0$.
For instance, in the $p$-spin spherical 
model, a whole branch of TAP solutions existing at zero temperature 
disappears as soon as we infinitesimally heat the system \cite{CGP1}
(see also appendix of \cite{FV}).


Looking at the other solution, we observed that
the BRST-susy  saddle point  does not yield a proper 
result either.
Indeed, it  counts TAP solutions that do not
 satisfy the Plefka criterion, 
i.e. solutions not corresponding to physical states. 
Furthermore, even its mathematical consistency is doubtful, since it is
 obtained setting $B=0$ (see appendices A and B), 
and, according to Plefka's analysis of the resolvent \cite{Ple0,Ple1}, 
this assumption is only justified if $x_{\rm P}>0$, 
while for $x_{\rm P}<0$
 a solution with 
$B\neq 0$ must be considered.  
It may happen that,
 upon passing to the quenched computation, the violation of the Plefka
 criterion  of the BRST solution may be cured. Certainly this happens at the Full-RSB 
lower band-edge, which, as we said above, is BRST supersymmetric \cite{BMY}.

In order to solve the problem of selecting a meaningful complexity
one can still try to generalize the 
proof of Kurchan \cite{K91} 
to arbitrary $u$, possibly using a procedure not involving 
a series expansion in $1/N$ and/or look for a quenched FRSB
solution  breaking the BRST-susy, thus solving 
a Parisi equation like the one of coupled replicas investigated 
in Ref. \cite{CR} but with a different boundary condition.
Both approaches are currently under investigation, 
together with the study of the quenched BRST solution\cite{CLPR}.

\vspace*{.5 cm}
{\bf Acknowledgments}: We thank A. Cavagna, I. Giardina and A. Annibale for many useful discussions.

\addcontentsline{toc}{section}{Appendix A}
\section*{Appendix A: The determinant of the Hessian of the TAP free energy.}
\label{app:A}

The inverse susceptibility matrix is
\BEA
&&\chi^{-1}_{ij}=-J_{ij}+a_i \delta_{ij}
\\
&& a_i=\frac{1}{\beta^2}\frac{1}{1-m_i^2}+\beta(1-q) 
+O\left(\frac{1}{N}\right)
\EEA
where the $J_{ij}$ are distributed according to 
\BEQ
P(J_{ij})=\sqrt{\frac{N}{2\pi}}\exp\left(-\frac{J_{ij}^2N}{2}\right)
\EEQ
As it is usually done,
we will not consider terms of order $1/N$ in the following computation.

The determinant can be written  with the help of
Grassmann variables, ($\eta$, ${\overline{\eta}}$), as
\BEQ
\det {\mathbf{\chi}}^{-1}=
\int\prod_{i=1}^N\d\eta_i~\d{\overline{\eta}}_i
\exp\left\{\sum_{ij}{\overline{\eta}}_i \chi_{ij}^{-1}\eta_j
\right\}
=\int\prod_{i=1}^N\d\eta_i~\d{\overline{\eta}}_i
\exp\left\{-\sum_{i<j}J_{ij}\left({\overline{\eta}}_i \eta_j
+{\overline{\eta}}_j \eta_i\right)
+\sum_i{\overline{\eta}}_i a_i \eta_i
\right\}
\EEQ

Its average over the disordered interaction is
\BEA
{\overline{\det {\mathbf{\chi}}^{-1}}}
&&=\int\prod_{i=1}^N\d\eta_i~\d{\overline{\eta}}_i
\exp\left\{-\frac{1}{2N}\left(\sum_{i}{\overline{\eta}}_i \eta_i\right)^2
+\sum_i{\overline{\eta}}_i a_i \eta_i
\right\}
\\
&&=\int_{-\infty}^\infty \frac{\d w}{\sqrt{2\pi/N}}
e^{-w^2N/2}
\int\prod_{i=1}^N\d\eta_i~\d{\overline{\eta}}_i
\exp\left\{
\sum_i{\overline{\eta}}_i \left(i w + a_i\right) \eta_i
\right\}
\nn
\\
&&=\int_{-\infty}^\infty \frac{\d w}{\sqrt{2\pi/N}}
\exp\left(-\frac{w^2N}{2}\right)
\exp\left\{
\sum_i \log \left(i w + a_i\right) 
\right\}
\nn
\\
&&=\int_{-\infty}^\infty \frac{\d w}{\sqrt{2\pi/N}}\exp\left(
N H(w)\right)
\nn
\\
H(w)&&\equiv -\frac{w^2}{2}+\frac{1}{N}\sum_i\log(iw+a_i)
\EEA

To compute the integral for large $N$ we make use of the saddle point approximation, thus evaluating the solution of
\BEA
&&\frac{\p H}{\p w}=-w +\frac{1}{N}\sum_i\frac{i}{iw+a_i}=0
\label{a1:wsp}
\\
&&\frac{\p^2 H}{\p w^2}=-1+\frac{1}{N}\sum_i\frac{1}{\left(iw+a_i\right)^2}<0
\label{a1:wcond}
\EEA
The second inequality is a condition that in most cases ensures that 
the integration path can be modified in order to cross the
 saddle point in the proper way, and is analogous to the maximum condition of the Laplace method. 
Changing $iw$ in
\BEQ
iv\equiv iw +\beta(1-q)=iw +a_i-\frac{1}{\beta}\frac{1}{1-m_i^2}
\EEQ
the stationarity condition for the saddle point reads
\BEQ
v\left[1-\frac{\beta^2}{N}\sum_i\frac{\left(1-m_i^2\right)^2}
{1+iv\beta\left(1-m_i^2\right)}
\right]=0
\label{a1:sp}
\EEQ
and the  condition for $H$ becomes:
\BEQ
1-\frac{\beta^2}{N}\sum_i\frac{\left(1-m_i^2\right)^2}
{\left[1+iv\beta\left(1-m_i^2\right)\right]^2}
>0
\label{a1:max}
\EEQ
The  saddle point equation (\ref{a1:sp}) has two solutions:
$v=0$, $v=v^\star\neq 0$.

\begin{itemize}
\item{Solution $v=0$.}

In this case
Eq. (\ref{a1:max}) simplifies in
\BEQ
1-\frac{\beta^2}{N}\sum_i\left(1-m_i^2\right)^2>0
\EEQ

This is exactly Plefka's criterion  characterizing a physically relevant 
solution.
The stationary value $v=0$ is a maximum of the exponent $H$
and corresponds to TAP solutions yielding the physical expression
of the linear susceptibility.

\item{Solution $v=v^\star$.}

For such a saddle point Eq. (\ref{a1:max}) can be written as 
\BEQ
1-\frac{\beta^2}{N}\sum_i\frac{\left(1-m_i^2\right)^2}
{1+iv\beta\left(1-m_i^2\right)}
+iv\frac{\beta}{N}\sum_i\frac{\left(1-m_i^2\right)^3}
{\left[1+iv\beta\left(1-m_i^2\right)\right]^2}
>0
\label{a1:max_vs}
\EEQ
of which the first two terms cancel each other.
Since $m_i^2<1$ always, this implies that in order to have
\BEQ
iv\frac{\beta}{N}\sum_i\frac{\left(1-m_i^2\right)^3}
{\left[1+iv\beta\left(1-m_i^2\right)\right]^2}
>0
\EEQ
must be $iv$ real and positive. 
In order to go back to BM notation
we define  the real variable $B=iv \beta$.
If $B>0$ the stationary point $B=i v^\star\beta$ is a maximum of $H$.
Inserting such a positive value into the saddle point Eq. (\ref{a1:sp}) 
one gets the inequality
\BEQ
1=\frac{\beta^2}{N}\sum_i\frac{\left(1-m_i^2\right)^2}
{1+B\left(1-m_i^2\right)}<
\frac{\beta^2}{N}\sum_i\left(1-m_i^2\right)^2
\EEQ
Thus violating Plefka's criterion.

\end{itemize}

Summarizing, if the Plefka criterion is satisfied the correct solution is $B=0$ while
 if it is
 not satisfied one must choose the solution with $B>0$. Therefore if one sets {\it a
 priori}
 $B=0$, then he must self-consistently check that the Plefka criterion is verified.
 Therefore,
 from this point of view,  the BRST solution  is mathematically inconsistent.
The Plefka criterion arises as a condition to determine which is the correct solution
 of the
 resolvent equation; the fact that it is also the condition for determine the correct 
saddle
 point for $B$, as we derived above, is not surprising, indeed the two computation are 
intimately related as we shown in the following appendix.


\addcontentsline{toc}{section}{Appendix B}
\section*{Appendix B: Identification of $\det {\mathbf{\chi}^{-1}}$
 Saddle Point Eq. and
Resolvent Eq. for $\mathbf{\chi}^{-1 }$}
\label{app:B}

In this appendix we would like to stress
the analogy between the saddle point equation for $w=-i[B/\beta+\beta(1-q)]$
(see App. A) and the equation for the resolvent  of the inverse
susceptibility. Using the notation of Ref. \cite{Ple1},
the resolvent is
\BEQ
R(z)=\frac{1}{N}\Tr\frac{1}{z-{\mathbf \chi}^{-1}}
=\frac{1}{N}\Tr\frac{1}{z-{\mathbf J}-a_i}
\label{res}
\EEQ
where the resolvent equation is
\BEQ
R(z)=\frac{1}{N}\sum_i\frac{1}{z-R(z)-a_i}
\label{reseq}
\EEQ
and the condition $\Im (R(z))>0$  must hold for $\Im (z)<0$.

We notice that  equation (\ref{reseq}) 
evaluated in $z=0$ is identical to Eq. (\ref{a1:wsp}) in App. A
provided the transformation $\Re (R(0))=iw$ is performed. Conversely the condition $\Im (R(z))>0$   for $\Im (z)<0$ is equivalent to (\ref{a1:wcond}). Indeed
the derivative of the resolvent comes out to be
\BEA
&&\frac{\d R}{\p z}=-\frac{Y(z)}{1-Y(z)}
\label{dRdz}
\\
&&Y(z)\equiv\frac{1}{N}\sum_i\frac{1}{\left(-z+R(z)+a_i\right)^2}
\EEA
The function $Y(z)$ is always positive, for any real $z$.

If we expand $R(z)$ around a given real value of $z=z_R$ for a small
negative imaginary part $-i \epsilon$
we get 
\BEQ
R(z_R-i\epsilon)=R(z_R)-i\eps~ Y(z_R)\left(1
-\left.\frac{\d R}{\d z}\right|_{z_R}
\right)
\EEQ
Thus, the condition on the imaginary part of R(z) for negative $\Im z$
(Pastur theorem \cite{Ple1,Pastur}) reads
\BEQ
\lim_{\Im z \to 0^-}\Im R(z)=\epsilon \lim_{\Im z \to 0^-} Y(z) 
\left(1-\frac{\d R}{\d z}\right)>0 \hspace*{ 1 cm}
\to
\hspace*{1cm}
\left.\frac{\d R}{\d z}\right|_{\Im z = 0}<1
\label{condPast}
\EEQ
Eq. (\ref{dRdz}) evaluated at $z=0$ satisfies condition (\ref{condPast})
if $Y(0)<1$.

If we set $R(0)=i w$ the equations (\ref{reseq})
 and (\ref{condPast}) evaluated 
at $z=0$ are equivalent respectively to Eqs. (\ref{a1:wsp})
and (\ref{a1:wcond}), thus legitimating this last equation as a validity 
condition for the saddle point of $H(w)$.

The resolvent equation has two  roots, which, for small $z$ and in the region
where $\beta^2\left(1-2 q +\left<m^4\right>\right)\simeq 1$, were evaluated 
e.g. in  Ref. \cite{Ple1}. 
They correspond to the $B=0$ and $B>0$ solutions of the previous section. 
The condition $\Im (R(z))>0$ for $\Im (z)<0$ selects one or the other
 solution depending on the value of $x_{\rm P}$ inasmuch as the condition 
(\ref{a1:wcond}) selects the correct solution $B=0$ or $B>0$ depending 
on the value of $x_{\rm P}$

\end{document}